# A Creepy World

**Didier Sornette and Peter Cauwels**
*Chair of Entrepreneurial Risk*
*Department of Management Technology and Economics*
*ETH Zurich*
*6 October 2013*

**Abstract:** Using the mechanics of creep in material sciences as a metaphor, we present a general framework to understand the evolution of financial, economic and social systems and to construct scenarios for the future. In a nutshell, highly non-linear out-of-equilibrium systems subjected to exogenous perturbations tend to exhibit a long phase of slow apparent stable evolution, which are nothing but slow maturations towards instabilities, failures and changes of regimes. With examples from history where a small event had a cataclysmic consequence, we propose a novel view of the current state of the world via the logical scenarios that derive, avoiding the traps of an illusionary stability and simple linear extrapolation. The endogenous scenarios are "muddling along", "managing through", and "blood red abyss". The exogenous scenarios are "painful adjustment" and "golden east".

**Table of contents**

### 1- How to boil a frog

When you put a frog in boiling water, the frog will jump out immediately. But, if you place her in a frying pan with cold water, and you increase the temperature very slowly, the frog will be unaware of the danger and will gradually adapt to its environment, until she is cooked to death. This phenomenon was under scientific investigation at the end of the 19$^{th}$ century. Zealous scientists argued about the optimal rate of temperature increase and the possible effect of the initial state and health of the poor animals. Anecdotal or not, the metaphor embraces a deep insight: the wisdom of how a slowly increasing load can remain unnoticed until the last small straw breaks the camel's back.

We all share creepy experiences in everyday life: a picture falling off the wall after it has been hanging there for years, a plastic shopping bag dropping off the table half an hour after you put it there. Things that seem to be stable can, all of a sudden fall, drop, fail, break, bend, snap or even die. It seems spooky. Is this magic or sheer bad luck? No one notices any change; no one sees it coming, until something big happens.

Even though these experiences may seem mysterious, there is always a scientific rationale. The surprise factor is just the consequence of our linear thinking, of the perception that all systems only gradually change, stuck in some kind of equilibrium. We naturally extrapolate such observations into the future, because we are not trained in seeing and understanding complex non-linear processes. But if you just sit down and take some distance, you will be able to enjoy the real beauty of our physical and social world and you will find out that hardly anything in life is really stable and in equilibrium, but rather slowly creeps towards a change of regime.

We will start our exploration with a description of the mechanics of creep. This is a textbook example, in material science, of a counter intuitive, non-linear, abrupt process. Creep will set the scene and become the thread throughout this piece. We will argue that such highly non-linear mechanisms, like creep, are not bound merely to the sphere of solid materials and physical systems, but also take place in our social, financial and economic fabric. This can be observed in the existence of tipping points, when a system is slowly pushed over the edge of a critical threshold and a runaway process sets in leading to a change of regime or bifurcation. We will back this assertion up with some examples from history, cases where a small event had a cataclysmic consequence. This will allow us to take a new, fresh look at the current state of the world. It will also give us a better sense of how we can construct future scenarios without falling in the trap of an illusionary stability and without using simple linear extrapolation.

### 2- Creepy mechanics

Creep, also known as "static fatigue", occurs when you put a material under a constant stress, which is below its mechanical strength, so that it does not break immediately. It is by waiting a sufficiently long time that the cumulative strain may finally end up in a catastrophic rupture [1].

The process is often divided into three regimes. During the primary regime, the stress actually strengthens the material, due to work hardening. This is well understood by clocksmiths who tap the teeth of brass gears and ratchets to give them a superior wear resistance. This net positive effect, however, fades out and at some point a balance is found between the strengthening due to work hardening and the weakening due to the occurrence of micro-fractures caused by the constant stress level. At that point, the secondary creep sets in. This stage is characterized by a quasi-constant imperceptible deformation rate. It seems as if the stress has no significant effect on the material; everything is stable, the setup looks robust. But over time, the system slowly evolves towards the tertiary creep regime. Suddenly, and unexpectedly, the strain rate accelerates and the material breaks. This is like a climbing rope that snaps. When a first fiber breaks, the load is distributed over the remaining fibers. Each time a new fiber snaps, the load on the other fibers increases. This creates a positive feedback mechanism that leads to a runaway process, the failure of the rope and the fall of the rock climber.

It is very important to stress here that the process we are describing is endogenous, which means that it is due to internal restructuration and self-organization. We are not talking about the failure of a material due to some kind of external event. Just picture a setup of a material under stress, no more and no less. For a very long time, the situation remains stable and apparently, everything looks perfectly fine. Then, all of a sudden, it breaks. This failure is deeply rooted in the fundamental properties of the system itself. The stable phase may hold on for years or centuries, but slowly, creep drives the system towards criticality. When a critical threshold is reached, something snaps and a new regime sets in.

Interestingly, in some materials, the catastrophic third creep regime is never reached. This is, for example, the case in materials that have a mix of solid and liquid-like properties. When a balance is found between creation and annihilation of micro-fractures, between damaging and healing forces, the situation can muddle through forever. An example is the very slow flow of window glass. It is a well-known fact that windows of churches, which were built several centuries ago, are thicker at the bottom than at the top due to creep flow of the "solid" glass, pulled down by gravity.

### 3- Peripeteia
This scientific curiosity may be *gefundenes Fressen* for physicists and mechanical engineers, but why would this be of any interest to investors, policy makers, regulators or financial analysts? There is a growing body of evidence that socio-economic systems behave in a similar way. More and more publications in the scientific literature [2-9] describe creep-like processes ending in abrupt rupture-like events occurring in our social, financial, ecological and economic fabric. From a mathematical perspective, this observation is not a surprise. It can be understood as a generic behavior of dynamical systems. According to general theorems of the theory of bifurcations and their classifications, there are only a finite number of ways by which systems can change regimes. Moreover, these changes of regime are not progressive but abrupt and present universal properties. This provides a meta-justification of the deep relationship between processes apparently as different as earthquakes, epileptic seizures and financial crashes, which is founded at a very fundamental level within the very structure of dynamical systems.

In such systems, stress may be exercised through control parameters such as investor trust, debt levels, deficits, food prices, the relative size of the real versus the financial economy, health, inequality, innovation, demographics, inflation, interest rates, global warming, bio-diversity, house prices and so on.

Before a critical threshold is reached, these parameters slowly drift, hardly changing the stress in the system, which has, at that point, a quasi-stable characteristic, only disturbed by noise. This gives us an illusion of control, we think that we understand the system and we assume that it will continue behaving as we expect it to for the indefinite future. We make naïve forecasts based on simple extrapolated trends without a real fundamental understanding of the underlying processes. Forecasters are happy and proud that they manage to find a trend and can fit a curve to it, without realizing that in fact they are blind. Examples are the credo that stocks always go up in the long run, the concept that inflation is contained because the velocity of money is low, the idea that governmental debt can always be refunded in the capital markets or that house prices always go up.

The real risks and opportunities in our modern, interconnected society can only be better understood if we get out of this fallacy. Only then will we be able to see and interpret the generic symptoms that occur when a system approaches criticality.

In a seminal paper in *Nature,* in 2009, Marten Scheffer, a Dutch environmental scientist, describes this as follows [10]:

*At first sight, it may seem surprising that disparate phenomena such as the collapse of an overharvested population and ancient climatic transitions could be indicated by similar signals. However… the dynamics of systems near a critical point have generic properties, regardless of differences in the details of each system.*

Extreme events are not unknowable unknowns. They are endogenous and thus deeply embedded in our social, economic, financial and ecological systems. As a consequence, there may be predictability. This can be, as is documented by Scheffer, in the form of *a critical slowing down of the system leading to an increase in autocorrelation in the resulting pattern of fluctuations* [10]. But there can also be early-warning signals in the form of "near misses" or "glitches" [11]. On the one hand, this is a positive message as it tells us that we can make a difference by observing, learning and taking action. But on the other hand, it also forces us to take responsibility because it prohibits us from hiding behind *force majeure* principles, simply blaming fate.

From the moment we wake up, till the time we go to sleep, massive amounts of information come our way. Our brain is hard-wired to deal with all that complexity by using simple, clear-cut, experience-based techniques, called heuristics, also known as rules of thumb. Simplification and pattern recognition help us through the day. It is in art, though, that our deep connectedness with turning points finds its strongest accepted expression. Sudden shifts are a natural part of our world and as such, we sense their aesthetic attraction. This is beautifully illustrated in the use of *peripeteia* in literature, which is, according to Aristotle, the most powerful part of a plot in a tragedy. *Peripeteia* is a sudden reversal or a turning point that emerges naturally from the circumstances. The example *par excellence* is the well-known story of Sophocles' Oedipus Rex [12], where the protagonist fulfills his destiny, to kill his father and marry his mother, actually by running away from it in the first place. The story takes the reader by the hand gently following its course. Then, something snaps. But, the turning point is not the result of some externality. Contrarily to a *Deus ex machina*, the turning point with *peripeteia* is logical within the frame of the story.

The aesthetics of Greek tragedies can help us escape our evolutionary determinism. This is a necessary condition to see the future challenges that our society is facing. The study of history can show us the way. By carefully scrutinizing the past beyond our heuristic reflexes and by disentangling *peripeteias* in past dramatic events, we may be able to construct more realistic scenarios for the future.

**4- The sand grains of history**
In his book *Civilization,* the British historian Niall Ferguson explains that civilizations are highly complex systems, made of a large number of interacting components organized more like a Namibian termite mound than an Egyptian pyramid. Interestingly, he writes [13]:

*Such systems can appear to operate quite stably for some time, apparently in equilibrium, in reality constantly adapting. But there comes a moment when they 'go critical'. A slight perturbation can set off a 'phase transition' from a benign equilibrium to a crisis – a single grain of sand causes an apparently stable sandcastle to fall in on itself.*

History is full of such sand grains, where a linear cause-and-consequence approach does not make any sense.

In Sarajevo, on June 28$^{th}$, 1914, a driver took a wrong turn. This triggered a sequence of events described most vividly by Mark Buchanan in his book *Ubiquity [3]:*

*The automobile stopped directly in front of a nineteen-year-old Bosnian Serb student, Gavrilo Princip. A member of the Serbian terrorist organization Black Hand, Princip couldn't believe his luck. Striding forward, he reached the carriage. He drew a small pistol from his pocket. Pointed it. Pulled the trigger twice. Within thirty minutes, the Austro-Hungarian Archduke Franz Ferdinand and his wife Sophie, the carriage's passengers, were dead. Within hours, the political fabric of Europe had begun to unravel...*
*When the First World War ended five years later, ten million lay dead. Europe fell into an uncomfortable quiet that lasted twenty years, and then the Second World War claimed another thirty million.*

More recently, on December 17, 2010, a Tunisian street vendor called Tarek al-Tayeb Mohamed Bouazizi set himself on fire in protest to the constant harassment and humiliations by a municipal official [14]. We could all witness, almost in real time, how this act was the catalyst for Tunisia's Jasmine Revolution and the following wider Arab Spring. In the end, rulers were ousted in Tunisia, Egypt, Libya and Yemen and civil uprisings and protests occurred throughout the whole Arab world.

Another remarkable case is given by Mikhail Gorbachev in a 2006 Op-ed piece for Project Syndicate [15]. According to the 1990 Nobel Peace Prize winner and last head of state of the Soviet Union, the nuclear meltdown at Chernobyl, even more than the launch of *perestroika*, was the real cause of the collapse of the Soviet Union five years later. He calls the event a *Turning Point,* mentioning that *there was the era before the disaster, and there is the very different era that has followed.*

During the decades before the accident, the Soviet system had slowly drifted towards criticality, with the loss of trust functioning in a similar way as the micro-fractures in a material subject to creep. In the presence of facilitating factors, such as a weakening leadership and the accumulation of managerial mistakes, growing nationalism and exploitation by the political elites to obtain power by independence from the USSR, the weakening of the communism ideology, Glasnost's policy and economic problems, the nuclear accident contributed to tip the system over the critical threshold. The nuclear chain reaction triggered a social and political chain reaction. The catastrophe was the direct consequence of a mal-functioning organization with a flawed safety culture, poor communication and a lack of trust in its employees and it led to a general disappointment in the competence of the communist leaders, a total failure of trust in the system and contributed to the final collapse of the Soviet Union.

The first question that comes to mind when studying historical events as well as physical processes is which control parameters apply. What factors have a similar effect on social systems that temperature, pressure, or stress have on a physical system? With our explanation of the collapse of the Soviet Union, we imply that the level of trust or more generally the social capital in a society may function in that way. Support for this assumption is documented in detail by Francis Fukuyama, who described in his book *Trust [16],* the impact that the *complicated and mysterious cultural process of social capital accumulation* has on different economic systems.

A more concrete and directly measurable example, though, is given in a 2011 paper by Marco Lagi and his colleagues at the New England Complex Systems Institute (NECSI) [17]. Their study shows that the timing of outbreaks of violent protests in North Africa and the Middle East coincided with large peaks in food prices. This is clearly demonstrated in Figure 1, which is taken from their paper. It shows the UN Food and Agriculture Organization (FAO) Food Price Index and the timing of reported food riots in recent years. Their observations suggest that high global food prices are a precipitating condition of social unrest in the World. Based on statistical analysis, they make the even stronger claim that food riots occur above a certain threshold of the index.

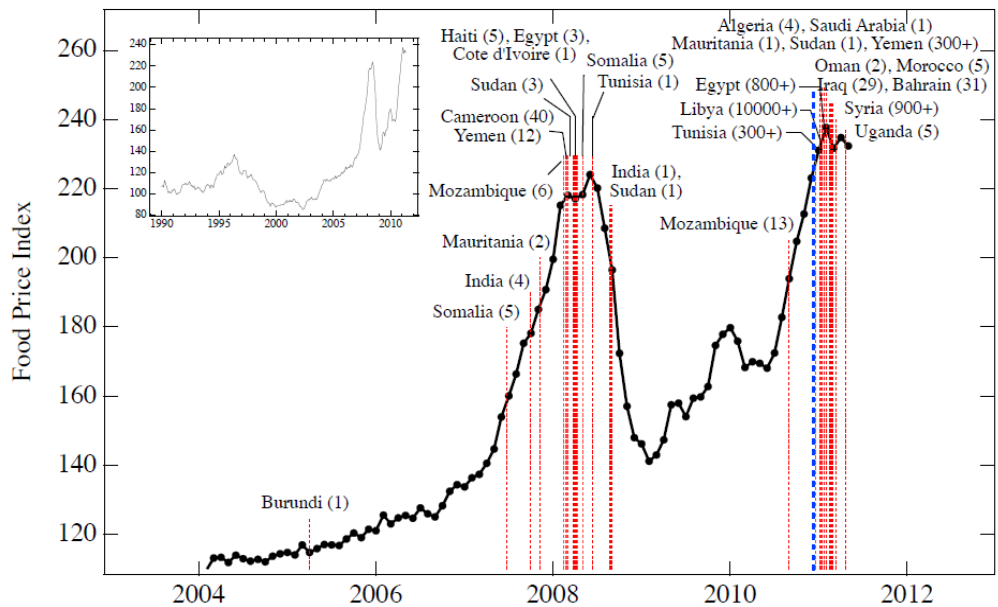

*Figure 1:* The FAO Food Price Index from January 2004 to May 2011. Red dashed vertical lines correspond to beginning dates of food riots and protests associated with the major recent unrest in North Africa and the Middle East. The overall death toll is given in parentheses. The inset shows the same price index from 1990 to 2011.The blue line corresponds to the date on which the authors of the report submitted their findings to the U.S. government [10].

The impact of food, in the form of scarcity due to drought and bad crops over several years, had a similar effect in France at the end of the 18$^{th}$ century. Adding to the mismanagement by the king and the nobility, the massive debt and sky-rocketing tax rates, it led to the French revolution in 1789, with sweeping consequences for Europe and beyond.

**5- A creepy economy**

Before the credit crisis of 2007-2008 pushed our economy into the great recession, economists, policy makers and central bankers, believed that we had transformed into a new world of continuous growth and prosperity. This was called the great moderation. It was a period of about two decades, characterized by robust and steady GDP growth, low and controlled inflation and unemployment and minimal financial volatility. Figure 2 gives some historical perspective. Economists were congratulating themselves, believing that those successes were the result of an improved economic policy making that was based on the scientific progress that their profession had made. It led Robert Lucas, a Nobel Prize winner of the University of Chicago, to declare in his 2003 presidential address to the American Economic Association that the *central problem of depression-prevention has been solved.* The great crash that followed shattered those dreams when a few hundred billion USD of losses in the financial sector cascaded into 5 trillion USD of losses in world GDP and almost 30 trillion USD of losses in global stock markets [18].

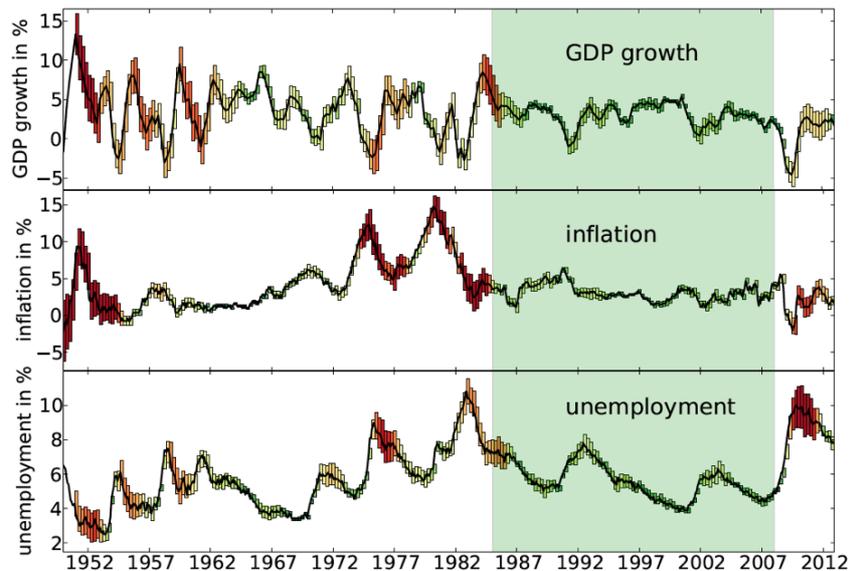

*Figure 2: The two decades of the great moderation, before the financial crisis in 2007-2008, led to the misguided belief of financial specialists that we had transformed into a new world of continuous growth and prosperity (taken from a talk on "How we can predict the next financial crisis", by Didier Sornette at TEDGlobal in 2013) [18].*

This rhetoric during the great moderation is a textbook example of what Reinhart and Rogoff call the perception that *this time is different* [19]. Financial professionals have a short memory and extreme events are quickly forgotten, even though they occur far more common than people seem to think. Moreover, as human beings, we tend to be driven by the hope or belief that the next innovation is always the one that really counts and will change the playground irreversibly, as none in the past has done before, hence the "new economy" mantra:

*Each time, society convinces itself that the current boom, unlike the many booms that preceded catastrophic collapses in the past, is built on sound fundamentals, structural reforms, technological innovation, and good policy [19].*

After studying eight centuries of data, Reinhart and Rogoff come to the conclusion that large-scale debt build-ups by governments, banks, corporations and consumers are the common theme to crises. Using our natural science lingo, we can say that their findings reveal that debt is an important control parameter of the economic and financial system. It drives the business cycle in a highly non-linear way with long periods of supposed stability punctuated with shocks and extreme events, basically flickering between the second and the third regime of the creep process, as we have described above.

In our previous piece for the Notenstein Academy, called *The Illusion of the Perpetual Money Machine* [20], we give a similar account of how, since the 1980s, our economy has seen a succession of bubbles and crashes. We paint a picture of a highly non-linear system with bursts of booms and busts driven by an extraordinary expansion of the financial sphere.

But, what drives this financial sphere? Using 150 years of data, Thomas Philippon of NYU Stern School of Business, uncovers the secrets behind the evolution of the U.S. Financial Industry. He comes to the conclusion that the share of finance in GDP increases when there is a need for intermediation from cash generating companies to cash absorbing companies.

In an illuminating paper [21], he explains how the growth of the U.S. Financial sector came in three waves followed by three crashes:

*The first large increase between 1880 and 1900 corresponds to the financing of railroads and early heavy industries… The second big increase between 1918 and 1933 corresponds to the financing of the Electricity revolution, as well as automobile and pharmaceutical companies …The third large increase, from 1980 to 2001, corresponds to the financing of the IT revolution. [21]*

The full historical picture can be found in Figure 3, which is taken directly from the paper.

We only need to put all the pieces of the puzzle together to come to a coherent view of the processes that actually drive our economy. Companies that want to commercialize or produce new technological products are generally new, cash poor, growth firms. So, in the process of innovation, financial intermediation is needed between old cash generating and new, innovative cash absorbing businesses. Consequently, the growth of the financial industry is deeply linked to the process of industrial innovation. As Bill Janeway explains in his book, *Doing Capitalism in the Innovation Economy [22]*, both speculation and innovation are two sides of the same coin, they mutually catalyze each other for growth and wealth creation. In other words, innovation in the "real" economy is catalyzed by innovation in the financial sphere.

Now, we can construct a universal scenario for a generic boom bust cycle. In a period of new innovations, like an industrial revolution, a boom sets in. But, industrial innovation needs financial innovation. Consequently, the size of the financial industry increases. This overshoots, however, and a disproportional growth of the banking industry finally causes a bust and puts an end to the cycle. This process can start again with a new wave of innovations.

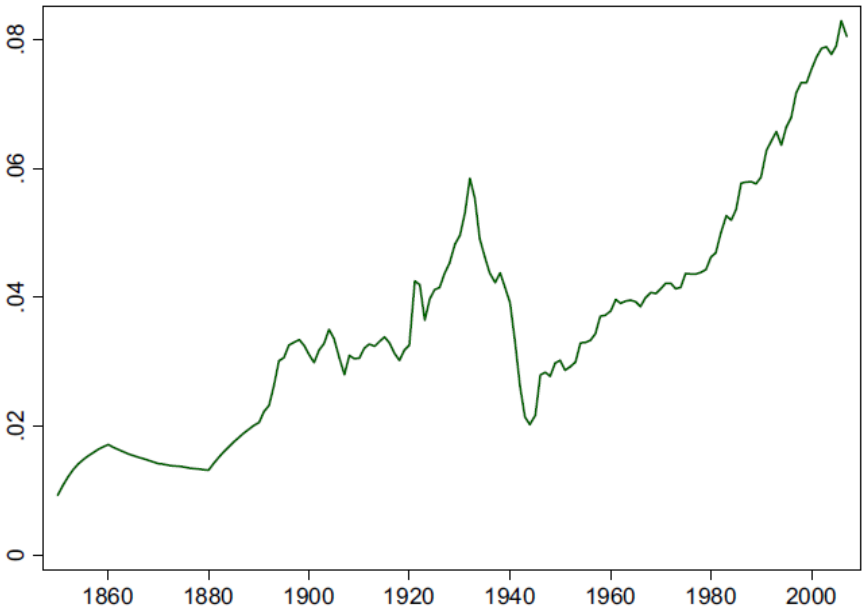

*Figure 3:* The share of the U.S. Financial Industry in the GDP. There are three large increases: 1880-1900, 1918-1933 and 1980-2001. Each of these is a period of great innovations where an intermediation from cash rich to cash poor companies was needed [21].

| P a g e    9

This is not just another description of the business cycle. We want to stress that the dynamics described here are highly non-linear and extremely asymmetric. Like in the mechanical process of creep, long periods of stability and pleasant growth are punctuated with extreme events, sudden shocks that reset the system. Because we are hard-wired to recognize simple patterns, we quickly forget the shocks and only see and extrapolate the local linear behavior of the system, forgetting that our economy is actually creepy.

**6- Predicting the present**
We have explained that history, in general, and economic systems specifically, behave in bursts, similar to the way creep precedes and foretells failures in material sciences. We argued that those bursts are not the result of some externality but that they are deeply rooted in the fundamental properties and resulting self-organization of the system itself. As there are no gods to blame, this puts the responsibility on us. This begs the need to understand our social systems better.

Before having the ambition to construct future scenarios, though, the first requisite is to have a better understanding of the present. This may seem a trivial thing to do. But, think about it: generally, we have no clue about the present instantaneous state of a social system such as an economy. We are using data that we get only monthly or quarterly and that comes with a lag, like unemployment or GDP, with a lot of incompleteness and imperfection. The metrics are outdated, static and do not fully reflect the true dynamical properties of the system, like correlation or volatility, and, the supposed underlying processes are no more than a theoretical abstraction, like the concept of equilibrium or the hypothesis of an efficient market. Let us now give examples of how new data, innovative metrics and dynamical models can be used to improve our understanding.

The act of feeling the pulse of our society was coined *predicting the present,* by Google researchers [23], who suggested that the use of Google Trends might give us a better understanding of the present. A nice practical example of this is the *Google Flu Trends monitor*. There is a close relationship between how many people search for flu-related topics and how many people actually have the flu [24]. As a consequence, by looking at the number of Google search queries, the present flu dynamics can be characterize much better than any other official record, beating standard approaches which usually lag by weeks. This can empower decision makers to react much more efficiently and adapt their policies in real time.

We decided to follow a similar approach to better understand how the world economic development and globalization unravels by asking what novel metrics would diagnose the rise of the East and the often-discussed decline of the West as well as quantify the pace of this transition. We imagined using an open source data set of nighttime light from the Defense Meteorological Satellite Program [25], with the nighttime lights on Earth taken as a proxy for economic development and energy consumption. Our study showed that, over the past two decades, the center of light of our planet recorded by different satellites operating between 1992 and 2009 shifted eastwards over a distance of roughly 1000 km, as illustrated in Figure 4. Moreover, the World center of light has moved in an accelerated fashion, echoing the economic boom of Asia. This figure, much more material and deeper analyses, can be found at the World at Night [26] webpage, which was developed for a master thesis project in our research group at ETH Zurich.

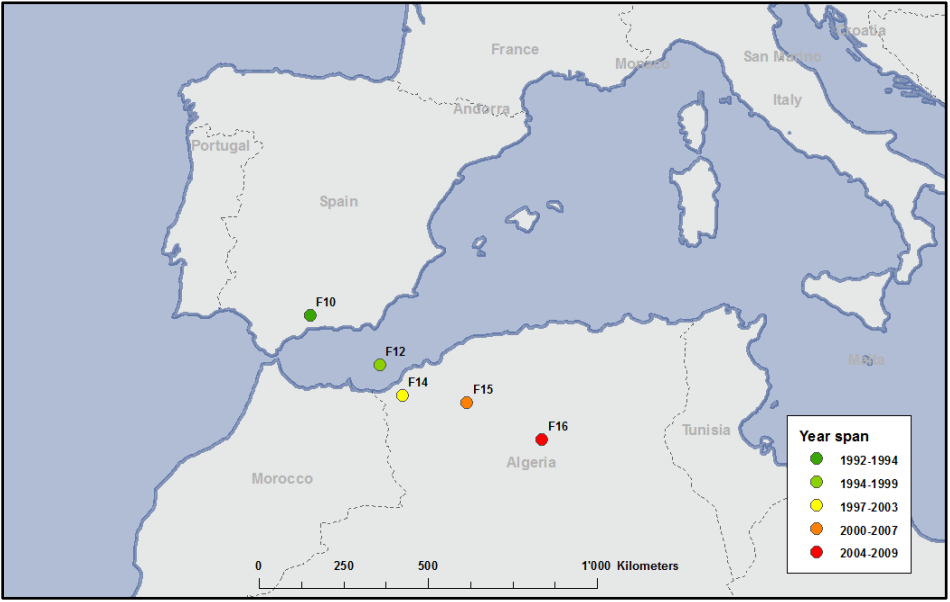

*Figure 4:* *The movement of the planetary center of light as recorded by 5 different satellites that operated between 1992 and 2009. Following a global economic regime shift, it can be seen moving eastwards over a distance of roughly 1000 km in that period of 17 years.*

Clearly, the availability of new data sources can improve our knowledge of the present. This is often called the *Big Data* revolution. There is a lot of new data out there, for example, in the form of Tweets or Google search queries, being analyzed by armies of researchers to enrich the current knowledge of our social systems. In financial markets, however, massive data flows have been available for quite some time. In fact, there is not so much need for more data but rather for adequate tools to mine it. The data is there, but we need to know where to look and how to interpret it. An example is the new *reflexivity index* that was developed by our research group at ETH Zurich [27,28]. This tool quantifies the level of endogeneity in financial markets, defined as the fraction of price-changes that are self-generated, like aftershocks from an earthquake, rather than the result of new, external or exogenous information or driving force. Our analyses show that the level of endogeneity in the S&P 500 index of U.S. equities has increased from 30% to approximately 75% in the last decade [27]; similar trends are observed in energy and commodities markets over that same period [28]. This means that markets are increasingly leading a life on their own, only partially driven by external news and are becoming more disconnected from the physical reality. One main reason is likely the rise of algorithmic trading and more specifically of High Frequency Trading (HFT).

Besides, there is increasing evidence of "glitches" in the system like the *flash crash* of 6 May 2010, the *hash crash* in 23 April 2013, and the recent three-hour halt on the Nasdaq Stock Market on 22 August 2013. Preliminary analyses suggest that these could be the outcomes of a system approaching criticality [27,28]. Clearly, financial markets have a very important function in our society, that of pricing, of providing efficient allocation of capital and of allowing risk diversification. But also here, the law of diminishing returns rules and, at a certain point, *more* does not necessarily mean *better*: more markets may push the system over into criticality.

Besides using good data and the right metrics, we also need to apply dynamical models that take into account non-linear behavior and the resulting non-stationarity of financial systems, in order to be able to calculate the risk of a critical transition. An example of such an analysis is given in Figure 5. We analyzed 60 years of quarterly data of Nominal GDP divided by the Total Liabilities of the Non Financial Sector in the U.S. This measure gives the efficiency of debt. It tells you how many USD of GDP are generated for each USD of debt in the Non Financial sector. It can be seen that, also here, the law of diminishing returns is at work; a 0.7 USD return on debt in 1953 dropped to 0.3 USD in 2013.

The question is whether this drop is sustainable. This can only be answered by using non-linear dynamical models. We analyzed the dataset using a proprietary model called *DS LPPL Trust$^{TM}$* that was developed by our research team at ETH Zurich. The model, whose engine is described for instance in [29,30], quantifies the criticality of the system; the risk for a change-in-regime. In Figure 5, the result of this analysis is given as a solid grey line. The results are alarming and point to a real risk for a critical transition. We explained that there is clear evidence that debt is a strong driver of our economic system. Our research shows that the effectiveness or usefulness of this debt is collapsing at an unsustainable rate. Moreover, it indicates that the point of a critical transition is nearby.

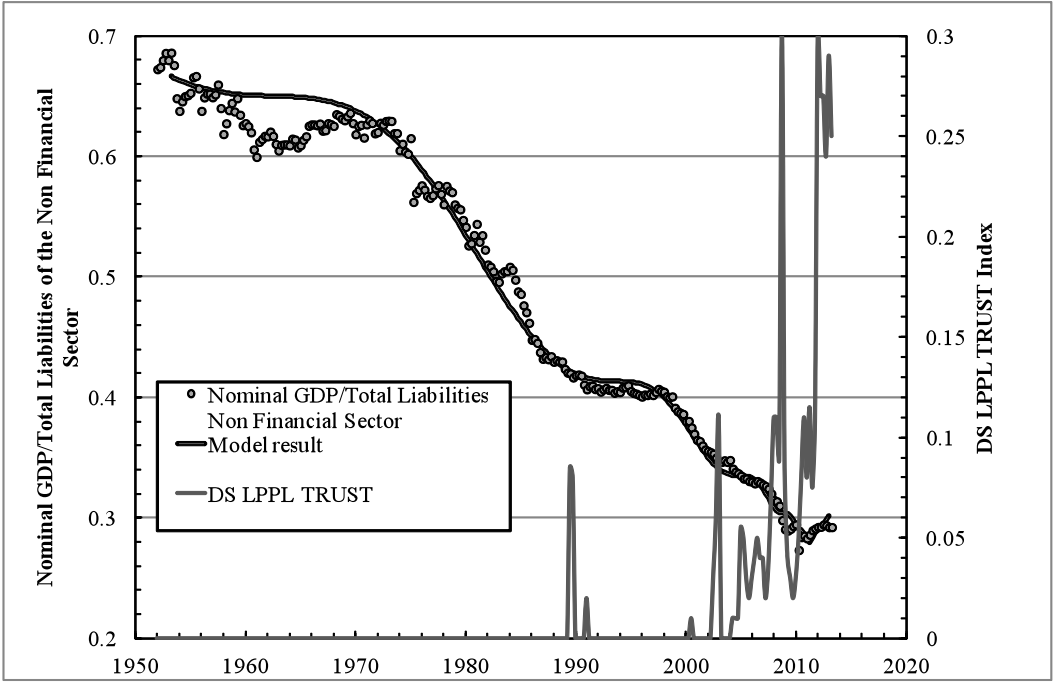

*Figure 5:* The efficiency of U.S. Non Financial debt. The dots tell you how many USD of GDP are generated for each USD of U.S. Non Financial debt. The black double line gives an LPPL model fit and the grey line gives the DS LPPL Trust index. When this index exceeds the 5% level, the process is not sustainable and there is a substantial risk for a critical transition of the system.

### 7- Scenarios for the future

Until now, we have argued that financial markets, economies and societies are controlled by slow and barely noticeable creep processes that often mature into abrupt nonlinear instabilities and changes of regime. This insight permits a natural classification of possible future scenarios making a distinction between those that are mainly endogenous and those that depend more on exogenous developments. We use the colorful terminology introduced by the executive board of Notenstein, as it matches well the classification that results from the creep-bifurcation theory.

**Endogenous scenarios:**

**"Muddling along"** is the scenario in which the system creeps under an existing set of stresses and constraints, where central banks and policy makers try to maintain a semblance of status quo, when pain slowly builds up with a general impoverishment of 99% of the population as a result of disguised inflation, confiscatory taxations and progressive dilution of social and retirement benefits. The science of creep and the mathematics of bifurcation both suggest however that, while "muddling along" can persist longer than often expected, it cannot be perennial and either evolves into a more stable regime ("managing through") or abruptly collapses into a system failure ("blood red abyss").

**"Managing through"** occurs when creep is counterbalanced by healing. While the forces of degradation are still present, restructuration and transformation can lead to adaptation and a kind of rebirth. This is exemplified by Canada, which extricated itself from a dire situation and exhibited a strong recovery in the 1990s, following courageous fiscal restructuration, deficit reductions, stringent inflation reduction and a reorientation of industrial activities to the novel economic realities. It should be noted however that Canada is a relatively small economy that benefitted from the strong growth of its giant southern neighbor and of a large part of the rest of the World, and thus could build a strong exportation base. In contrast, the task for Europe, as well as the U.S. and Japan, seems herculean due to their size and impact, as the economic growth in the world and in particular in Asia is expected to considerably slow down and a currency war-of-sort is unfolding in beggar-thy-neighbor attempts.

**"Blood red abyss"** is the straightforward, likely and very painful final stage of creep, the tertiary regime of sudden accelerated damage, ending in the failure of existing institutions. There are several pathways to the "blood red abyss''. We can imagine a complete finance freeze, in which, one day, market participants refuse to continue to play along with central banks and, through contagion, their trust in sovereign debts evaporates abruptly in a massive and generalized effective bank run. Or, the policy makers who implement their "muddling along" tactics engineer unwillingly waves of social upheaval powered by the disenchanted and frustrated unemployed youths, which spiral into constitutional crises, perhaps even revolutions and wars.

**Exogenous scenarios:**

**"Painful adjustment"** occurs under the influence of external factors and forces. In the material science analogy, it can take the form of a repair provided by an externally applied plaster or glue. It is the response to exogenous conditions and influences that constrains the economy to change its trajectory, to adapt its internal organization, most likely unwillingly and with resistance and hesitations. In all likelihood, it is not more stable than "muddling along" but buys time and, to those unaware of the present piece, gives the illusion of a long-term solution. In our previous analysis [13], we diagnosed without ambiguity that much of the developed world has inherited a state resulting from three decades of unfunded consumption and of living beyond one's mean, which we called "the Illusion of the Perpetual Money Machine". The needed readjustments are obviously severe and of large impact, from deep fiscal reforms, the reorganization of a sustainable social security, more thinking about how to ensure stronger social capital and better distribution of wealth, which is the glue between people at the origin of the health of nations, redesigning incentives at all levels of society and fighting inequity not with easy access to credit to increase consumption but with improved participation, e.g. by emphasizing excellent targeted education and continuous education (without e.g. the massive student loans debt).

Before innovation and growth take again control towards a better health of the economy, the readjustment will necessarily lead to painful destruction of buying power and entitlements. Readjustments may be triggered or further necessitated by other exogenous conditions, such as more severe adverse impacts from global climate change, large-scale environmental degradations, 1918-like flu pandemics and others. While we, humans, are obsessed with our human-made problems and calamities, we tend to forget that we live in and from a "small planet", our still unique spacecraft for survival in the immensity of a very inhospitable and immense space.

**"Golden East"** is the exogenous scenario where the solution to worsening economic conditions in the West could come at least in part from Chinese and Eurasian saviors. Without endogenous healing, the structure continues to hold because external support is added and the applied stress changes in nature. With the emergence of a large middle class in China, India and Indonesia, the growth of demand in eastern markets could foster exports of developed economies to their new customers in need of high-tech and high fashion products. Moreover, Europe could become even more the touristic attraction of the world of wealthy Americans, Asians and Africans (with the risk of having tourism as a "resource curse"), offering its historical cities, deep-rooted cultures, arts and savoir-vivre. Without doubt, this would also require a significant reorganization. However, given the exogenous dependence on the East, this scenario does not appear stable either and would likely unfold sooner or later in one of the four previous ones, if only because the impact would also likely unfold on the global ecological footprint as suggested by the dismal state of the environment in growth-hungry China.